\documentclass[onecolumn,12pt]{IEEETran}
\usepackage{bbm}
\usepackage{amsfonts}
\usepackage{tipa}
\usepackage{amssymb}
\usepackage{mathrsfs}
\usepackage[lined,boxed,commentsnumbered]{algorithm2e}
\usepackage{graphicx}
\usepackage{amsfonts,amssymb,amsmath}
\usepackage{latexsym}
\usepackage{color, soul}
\usepackage{multirow}
\usepackage{hyperref}
\usepackage{epsfig,epstopdf}
\usepackage{array}

\newtheorem{thm}{Theorem}

\begin{document}
\title{
Fundamental Limits of Caching: Improved Bounds For Small Buffer Users}
\author{Zhi~Chen \IEEEmembership{Member,~IEEE}
        Pingyi~Fan~\IEEEmembership{Senior Member,~IEEE}
        and~Khaled~Ben~Letaief~\IEEEmembership{Fellow,~IEEE}
\thanks{Z. Chen is with the Department of Electrical and Computer Engineering, University of Waterloo, Waterloo, Ontario, Canada, N2L3G1, (email: z335chen@uwaterloo.ca).
P. Fan are with the Department of Electrical Engineering, Tsinghua University, Beijing, China, 100084 (email: fpy@tsinghua.edu.cn).
K. B. Letaief is with Department of Electronic and Computer Engineering,
Hong Kong University of Science and Technology, Hong Kong (e-mail:
eekhaled@ece.ust.hk).
}}

\maketitle

\maketitle
\baselineskip 24pt
\begin{abstract}\\
\baselineskip=18pt
In this work, the peak rate of the caching problem is investigated, under the scenario that the users are with small buffer sizes and the number of users is no less than the amount of files in the server. A novel coded caching strategy is proposed for such a scenario, leading to a lower peak rate compared to recent results in the literature. Furthermore, it is verified that our peak rates coincides with the cut-set bound analytically in an information-theoretic view.
\end{abstract}
\IEEEpeerreviewmaketitle

\section{Introduction}
Caching, a technique playing a crucial role in combatting the peak hour network traffic congestion, receives increasing attention recently. A natural way to reduce peak hour traffic is to duplicate some contents at the end users. In the literature, there are several works focusing on investigating how to duplicate fractions of files at end users so that the peak rate is minimized and network congestion is reduced. Usually, caching works in two phases. One is the placement phase, which is performed during off-peak times. The other is the delivery phase, performed during rush hours when network resources are scarce. The general model with caching strategy were discussed in \cite{dowdy1982file}-\cite{kol2011demand} where no coding strategy was applied and the gain comes only from local duplication. However, if each user is equipped with a cache with a small size compared with the amount of the content in the server, this gain is readily observed to be negligible.

In \cite{kol2011index}, the index coding strategy was discussed. In \cite{niesen2014fundamental}, a new coded caching strategy from an information-theoretic perspective was proposed to achieve a new achievable rate region for general scenarios, where some finite rate-cache pairs were firstly derived and then the lower convex envelope of these points is shown to be achievable by memory sharing. This strategy was shown to enjoy both the local gain from duplication as well as the global gain from coding.
This fundamental idea was then extended to \cite{niesen2014decentralized} where a decentralized coded caching algorithm was presented and to \cite{niesen2014nonuniform} where the non-uniform demand scenario was investigated. In \cite{clancy2014secure}, the secure issue with coded caching was investigated.

In this work, however, we investigate the fundamental achievable rate for a special case where all users are equipped with a cache of a small size. In this case, appropriate coded duplication of contents is essential to reduce the delivery rates.
To this end, we introduce a new coded caching strategy and it is shown that the rate of this strategy coincides with the lower cut-set bound when the cache size is rather small. With memory sharing, it is shown that our strategy outperforms the strategy proposed in \cite{niesen2014fundamental} in terms of achievable delivery rates when the cache size is relatively small.

\section{Problem Setting}
A system consisting of one server and $K$ users is considered. An error-free link is assumed to be shared by all users connecting the server, where $N$ files are stored for fetching. We also assume that each user is equipped with a cache of size $Z_k$ ($k=1, \ldots, K $) and each user is assumed to request only one full file. The aim is to design a novel coded strategy to achieve a lower peak rate that can guarantee each user obtaining the file requested, compared with the recent results on caching problems in \cite{niesen2014fundamental}. In this work, we turn our interest on the special case that all users are with small buffer sizes ($Z_k \le 1/K $) and $K \ge N$, i.e., the amount of users is no smaller than that of the files in the server.

For clarity, we denote the smallest peak rate achieved by our strategy
by $R(M)$, i.e., the cache-rate pair ($M$,$R(M)$) is on the boundary of the achievable region, where $M$ denotes the cache size of all users. For comparison,
we denote the minimum peak rate achieved in
\cite{niesen2014fundamental} by $R_c(M)$
and the lower cut-set bound by $R^*(M)$.

\section{Main Results}
\begin{thm} \label{thm:1}
For $N \in \mathbb{N}$ files and $K$ ($K = N$) users each with cache of size $M=1/N$,
the cache-rate pair ($1/N,N-1$) is achievable. Furthermore, if $M \in [0,1/N]$,
\begin{align}
R(M) \le N(1-M)
\end{align}
is achievable.
\end{thm}

\begin{thm} \label{thm:2}
For $N \in \mathbb{N}$ files and $K$ ($K \in \mathbb{N}$ and $K>N$) users each with cache of size $M=1/K$,
the cache-rate pair ($1/K,N-N/K$) is achievable. Furthermore, if $M \in [0,1/K]$,
\begin{align}
R(M) \le N(1-M)
\end{align}
is achievable.
\end{thm}

\begin{thm} \label{thm:3}
For $N \in \mathbb{N}$ files and $K$ ($K=N$) users each with cache of size $M \le 1/N$, the achievable rate coincides with the associated cut-set bound.
\end{thm}

\begin{thm} \label{thm:4}
For $N \in \mathbb{N}$ files and $K$ ($K \in \mathbb{N}$ and $K>N$) users each with cache of size $M \le 1/K$, the achievable rate coincides with the associated cut-set bound.
\end{thm}

Note that in \cite{niesen2014fundamental}, the achievable rate with
$M=1/K$ is on the line connecting the two cache-rate pair points ($0,N$) and the first non-trivial point ($N/K, \min\left((K-1)/2,N(K-1)/K \right)$)\footnote{Note that in \cite{niesen2014fundamental} only rates of a number of points with cache size of $tN/K$ ($t=0,\ldots,N$) are directly derived and then the achievable cache-rate region is determined by the lower convex envelope of these points. It is readily observed that the non-trivial direct-derived achievable point with the smallest cache size is hence the point with cache size $N/K$.} and is hence given by,
\begin{align}
R_c(\frac{1}{K})= &\frac{\min \left(\frac{K(K-1)}{2},N(K-1) \right)-KN}{N} \cdot \frac{1}{K}+N  \label{eq:proof:compare_0} \\
&=N-1 + \min(\frac{K-1}{2N},1-\frac{1}{K}) \label{eq:proof:compare_1}\\
&\geq N-1 + \min(\frac{N-1}{2N},1-\frac{1}{N}) \label{eq:proof:compare_2}\\
&= N - \max \left( \frac{N+1}{2N}, 1/N \right) \\
&\ge N-\frac{K}{N} =R(\frac{1}{K}). \label{eq:proof:compare_3}
\end{align}
where the inequalities in (\ref{eq:proof:compare_2}) and (\ref{eq:proof:compare_3}) follow from the setting that $K \ge N \ge 1$.
Note also that the inequality in (\ref{eq:proof:compare_3}) strictly holds as long as $N>1$, which demonstrates the gain achieved by our coding strategy over the strategy designed in \cite{niesen2014fundamental} for the small cache size scenario.

Furthermore, with our coding strategy, we have
\begin{align}
R(1/K)&=N(1-1/K) \\
&= \min\left(\frac{K-1}{2},\frac{N(K-1)}{K} \right), \quad \mbox{if $K \ge 2N$}\nonumber\\
&=R_c(N/K), \quad \mbox{if $K \ge 2N$.} \nonumber
\end{align}
Which is an encouraging result. In other words, with a smaller cache size $M=1/K$, the designed coding strategy can achieve a rate no smaller than that in \cite{niesen2014fundamental} with the cache size $M=N/K$ if $K \ge 2N$.

Therefore, compared with \cite{niesen2014fundamental}, the rate with the
cache size of $M<N/K$ is improved by our results through memory sharing,
where the exact expression of the achievable rate with $0 \leq M \leq N/K$ is given on top of next page.



\section{Examples}

{\bf Example 1.} In this example, we set $N=K=3$, i.e., a system consists of three files in the server and three intended users. Let $W_1=A$, $W_2=B$ and $W_3=C$. We would like to show that
the $(M,R)$ pair ($1/3, 2$) is achievable.

With cache size $M=1/3$, we split each file into three subfiles with equal size, i.e., $A=(A_1,A_2,A_3)$, $B=(B_1,B_2,B_3)$ and $C=(C_1,C_2,C_3)$.
In the placement phase, the cache content of user $k$ is designed to be $Z_k=(A_k \oplus B_k \oplus C_k)$, which is an XORed version of three subfiles from different files in the server.

In the delivery phase, let us consider an example that user 1 requires $A$, user $2$ requires $B$ an user $3$ requires $C$. Hence, to obtain
the missing files for user 1, we should transmit $B_1$ and $C_1$ to obtain $A_1$ from the XORed subfile in $Z_1$ as well as $A_2$ and $A_3$ for the missing files of
$A$. In a similar manner, for user 2 requesting file $B$, the server need to transmit $B_3$ for the missing part of $B$ ($B_1$ is obtained from the shared link satisfying user 1). In addition, the server transmits
$C_2$ to obtain $B_2$ (as $A_2$ has been transmitted and received by user 2).
Note that the server has satisfied user 3 since the missing subfiles
$C_1$ and $C_2$ are already received by it. In addition, with the received
$A_3$ and $B_3$ from the shared link user $3$ can obtain $C_3$ from the cached $A_3 \oplus B_3 \oplus C_3$.

Therefore, the server has to transmit ($B_1$, $C_1$, $A_2$, $A_3$, $B_3$, $C_2$) to satisfy the requests of all users in this example. In a similar manner, all other requests can be satisfied.
Since each subfile has rate $1/3$, the total rate $2$ is achievable.

On the other hand, the cut set bound derived in \cite{niesen2014fundamental} indicates the minimum rate is $R^*(1/3)=3-3/3=2$ and is identical to the achievable rate. By cache sharing, we conclude that the achievable rate coincides with the cut set bound if $0 \leq M \leq 1/N$.



{\bf Example 2.} In this example, we consider a system with a server of $4$ files and $4$ users, i.e., $N=K=4$. The four files are termed as $W_1=A$, $W_2=B$, $W_3=C$ and $W_4=D$.

Consider the case with the cache size $M=1/4$. In this example, we split each file into four parts of equal size, i.e., $A=(A_1,A_2,A_3,A_4)$, $B=(B_1,B_2,B_3,B_4)$, $C=(C_1,C_2,C_3,C_4)$ and $D=(D_1,D_2,D_3,D_4)$.
In the placement phase, we let user $i$ caches the XORed subfile $Z_k=(A_k \oplus B_k \oplus C_k \oplus D_k)$.

In the delivery phase, for instance, consider that user $i$ requires $W_i$, i.e., user 1 requests A, user 2 requests B, user 3 requests C and user 4 requests D. We can satisfy all requests of different users by sending
($A_2$, $A_3$, $A_4$, $B_1$, $B_3$, $B_4$, $C_1$, $C_2$, $C_4$, $D_1$, $D_2$, $D_3$). It is observed that with this transmission subfile list, all missing subfiles can be received by intended users. In addition, it is readily verified that the intended subfile which is XORed in the cache of each user is also obtained by XORing the three other XORed subfiles. For example, for user 1, it receives $B_1$, $C_1$ and $D_1$, hence $A_1$ is also fetched by
$(A_1 \oplus B_1 \oplus C_1 \oplus D_1) \oplus B_1 \oplus C_1 \oplus D_1$. In a similar manner, user 2, user 3 and user 4 can also obtain $B_2$, $C_3$ and $D_4$ respectively. Therefore, by sending these subfiles, all user requests are satisfied with rate $3$, as the rate of each subfile is $1/4$.

Similarly, we can realize any possible requests with rate $3$ with the cache size $M=1/4$. Hence, the cache-rate pair ($1/4$, $3$) is achievable and can be verified to coincide with the cut-set bound, which is $R^*(1/4)=4-4 \cdot 1/4=3$. Therefore the cut-set bound is achievable if $0\leq M \leq 1/4$.

Akin to \textbf{Example 1 and 2}, the cache-rate pair ($1/N$,$N-1$) is achievable for an arbitrary number of files $N$ in the server with the same number of users as that of the files in the server, i.e., $K=N$. The proof for this general case is left to the next section.

{\bf Example 3.} Consider a system with $N=3$ files and $K=4$ users.
We term each file as $W_1=A$, $W_2=B$ and $W_3=C$.
Consider the case with cache size $M=1/4$. We split each file into 12 parts of equal size, i.e., $A=(A_1,\cdots,A_{12})$, $B=(B_1,\cdots,B_{12})$ and $C=(C_1,\cdots,C_{12})$. Each cache can therefore store three subfiles.
In the placement phase, we let user $i$ caches the three XORed subfiles as
$$Z_i=(A_{3(i-1)+j} \oplus B_{3(i-1)+j} \oplus C_{3(i-1)+j}), \quad j=1,2,3.$$
Hence one user caches 9 exclusive subfiles in an XORed version and any subfiles partitioned in the server can be found in the cache of one and only one user.

In the transmissions phase, let us assume that user $1$ needs $A$, user $2$ needs $B$, user $3$ needs $C$ and user $4$ needs $A$.
To fully exploit the coded caching strategy, we then delivery the subfiles ($B_1$, $C_1$, $B_2$, $C_2$, $B_3$, $C_3$) for user 1 to XOR $A_1$, $A_2$ and $A_3$.
By delivering of these subfiles, $B_1$, $B_2$ are received by user $2$ and $C_1$, $C_2$ are received by user $3$. Similarly, we deliver ($A_4$, $C_4$, $A_5$, $C_5$, $A_6$, $C_6$) for user 2 to obtain $B_4$, $B_5$ and $B_6$.
($A_7$, $B_7$, $A_8$, $B_8$, $A_9$, $B_9$) for user 3 to obtain $C_7$, $C_8$ and $C_9$. ($B_{10}$, $C_{10}$, $B_{11}$, $C_{11}$, $B_{12}$, $C_{12}$) for user 4 to obtain $A_{10}$, $A_{11}$ and $A_{12}$.

Hence, by delivering these $24$ subfiles, user $2$ receive the complete file $B$ and user $3$ receive the entire file $C$. However, user 1 still lacks the subfiles ($A_{10}$, $A_{11}$, $A_{12}$) and user 4 is in need of the subfiles ($A_1$, $A_2$, $A_3$). To exploit the side information at the caches, we hence delivery ($A_1 \oplus A_{10}$, $A_2 \oplus A_{11}$ and $A_3 \oplus A_{12}$). By doing so, we can fulfil the requests of all users with delivery of 27 subfiles, i.e., rate $R(1/4)=27/12=9/4$ is achievable for this case. Similarly, it can be readily shown that this rate is achievable for any other possible requests.

It is worth pointing out that, the cut-set bound at the point $M=1/4$ is $R^*(1/4)=3-3/4=9/4$ and identical to the achievable rate $R(1/4)$. Thanks to cache sharing, the cut-set bound is therefore achievable in the interval $M \in [0,1/4]$ in this example.

{\bf Example 4}. Consider the case of a server with $3$ files and 5 users.
We term each file as $W_1=A$, $W_2=B$ and $W_3=C$.
Consider the case with cache size $M=1/5$. We split each file into $3 \times 5=15$ parts of equal size, i.e., $A=(A_1,\cdots,A_{15})$, $B=(B_1,\cdots,B_{15})$ and $C=(C_1,\cdots,C_{15})$ and each cache can store three subfiles.
In the placement phase, we let user $i$ caches the three XORed subfiles as
$$Z_i=(A_{3(i-1)+j} \oplus B_{3(i-1)+j} \oplus C_{3(i-1)+j}), \quad j=1,2,3.$$
Each user then stores $9$ exclusive subfiles in an XORed version and each subfile can be found
in the cache of one and only one user.


In the transmissions phase, let us assume that user $1$ needs $A$, user $2$ needs $B$, user $3$ needs $C$, user $4$ needs $A$ and user $5$ requests $B$.
Similar to {\bf Example 3}, we deliver the subfile list ($B_1$, $C_1$, $B_2$, $C_2$, $B_3$, $C_3$) for user 1 to XOR $A_1$, $A_2$ and $A_3$,
Therefore $B_1$, $B_2$ are received by user $2$ and user $5$,  while $C_1$ and $C_2$ are received by user 3. Similarly, we deliver ($A_4$, $C_4$, $A_5$, $C_5$, $A_6$, $C_6$) for user 2 to obtain $B_4$, $B_5$ and $B_6$;
($A_7$, $B_7$, $A_8$, $B_8$, $A_9$, $B_9$) for user 3 to obtain $C_7$, $C_8$ and $C_9$; ($B_{10}$, $C_{10}$, $B_{11}$, $C_{11}$, $B_{12}$, $C_{12}$) for user 4 to obtain $A_{10}$, $A_{11}$ and $A_{12}$; ($A_{13}$, $C_{13}$, $A_{14}$, $C_{14}$, $A_{15}$, $C_{15}$) for user $5$ to obtain $B_{13}$, $B_{14}$ and $B_{15}$.

Hence, by delivering these $30$ subfiles, user $3$ receive the entire file $C$. However, user $1$ still requests the subfiles ($A_{10}$, $A_{11}$, $A_{12}$), user $2$ requests ($B_{10}$, $B_{11}$, $B_{12}$), user $4$ requests ($A_1$, $A_2$, $A_3$) and user $5$ requests ($B_{4}$, $B_{5}$, $B_{6}$).
To exploit the side information at the caches, we can delivery the XORed version of the subfiles, i.e., ($A_1 \oplus A_{10}$, $A_2 \oplus A_{11}$, $A_3 \oplus A_{12}$, $B_{11} \oplus B_{5}$ and $B_{12} \oplus B_{6}$). With this coded transmission, all intended users can completely obtain the subfiles requested.
We therefore fulfil the requests of all users by delivery of only 36 subfiles, i.e., rate $R(1/5)=36/15=12/5$ is achievable for this case.
In a similar manner, it can be readily shown that this rate is achievable for any possible requests.

It is worth pointing out that, the cut-set bound at the point $M=1/5$ is $R^*(1/5)=3-3/5=12/5$ and equals the achievable rate $R(1/5)$. By memory sharing, the cut-set bound is therefore achievable in the interval $M \in [0,1/5]$ in this example.

\section{Proof Of Theorems }
We now present the achievable scheme for an arbitrary number of users with $K \ge N$. We shall show that with the cache size of $M \leq 1/\max(N,K)$, the delivery rates presented in Theorem \ref{thm:1}-\ref{thm:2} are achievable and the cut-set bound is met for such points with cache size $M \leq 1/\max(N,K)$.

\subsection{Proof of Theorem 1}
Here we prove Theorem 1 for the case with an equal number of files and users, i.e., $N=K$. We prove it in two folds. Firstly, we verify that the point ($1/N$,$N-1$) is achievable by a constructed coded caching scheme. Secondly, we show that any points with $M<1/N$ can achieve a rate of $N-NM$ by memory sharing.

Let us define the files as $W_i$ ($i=1,\ldots,N$) and split each file into $N$ subfiles, i.e., $W_i=(W_{i1}, \ldots, W_{iN})$. In the placement phase, the cache of user $j$ is designed to be $Z_j = W_{1j} \otimes \ldots \otimes W_{Nj}$, an XORed version of subfiles, which contains one and only one subfile from all files. With this coded placement scheme, each user caches some exclusive part of all files.

In the delivery phase, if the users request $L \leq N-1$ files, we can simply transmit these requested files and the delivery rate is $L$ files. We then move to the case that the users request $N$ files, i.e., each user requests a different file. Due to symmetry, we only need to study the case that user $i$ requests file $W_i$. The transmission algorithm is therefore presented as follows.
\begin{itemize}
\item For the first file, we transmit the subfiles $W_{12}$, $\ldots$, $W_{1N}$.
\item For the $i$th ($1<i<N$) file, we transmit the subfiles $W_{i,1}$, $\ldots$, $W_{i,i-1}$, $W_{i,i+1}$, $\ldots$, $W_{1N}$.
\item For the $N$th file, we transmit the subfiles $W_{N1}$, $\ldots$, $W_{N,N-1}$.
\end{itemize}
As for each file $(N-1)/N$ fraction of it is delivered, we totally deliver $N-1$ files.

With this transmission, we argue that each user can obtain the files requested. For instance, for the $i$th user requesting $W_i$, it can obtain all subfiles except $W_{ii}$ from the delivery of $W_i$ directly. In addition, user $i$ receives
all $W_{ki}$ ($j \neq i$) subfiles from file $W_k$. Hence it can obtain the subfile $W_{ii}$ by
\begin{align}
W_{ii}=&(W_{1i} \oplus \ldots \oplus W_{Ni}) \oplus W_{1i} \oplus \ldots \oplus W_{i-1,i} \nonumber \\
&\oplus W_{i+1,i} \oplus \ldots \oplus W_{Ni} \\
=&W_{ii} \nonumber
\end{align}
Therefore, user $i$ can obtain all subfiles of $W_i$ and construct the complete file $W_i$. In a similar manner, all users can obtain the complete file requested and the cache-rate pair $(1/N,N-1)$ is hence achievable for this special case. Moreover, due to symmetry, we can conclude that the cache-rate pair $(1/N,N-1)$ is achievable for all possible requests.

On the other hand, with the two achievable points, i.e., ($0,N$) and ($1/N$,$N-1$) taken into account, we can achieve a rate of $R(M)=N(1-M)$ for the cache size $0 \le M \leq 1/N $ by memory sharing. Theorem 1 is hence proved.

\subsection{Proof of Theorem 2}
Here we prove Theorem 2 for the case with $N<K$.
The files are defined by $W_i$ ($i=1,\ldots,N$) and we split each file into $NK$ subfiles, i.e., $W_i=(W_{i,1}, \ldots, W_{i,NK})$.

In the placement phase, the cache of user $i$ is designed to
store $N$ XORed version of subfiles, which are,
$$Z_i=W_{1,N(i-1)+j} \oplus \cdots \oplus W_{N,N(i-1)+j}, \quad j=1, \ldots, N.$$
With this coded placement scheme, each user caches some exclusive part of all files and the union set of the caches comprises all $N$ files in the server.

In the delivery phase, if all users request $L$ ($L \leq N-1$) distinct files in total, we can simply transmit these requested files one by one and the total amount of files delivered is $L$ files and the associated rate is less than $N-N/K$. We then move to the case that all $N$ files are requested. Suppose
user $i$ requests the file $W_{d_i}$ and correspondingly the subfile $W_{i}$ is requested by totally $k_{i}$ users. By definition, we hence have $\sum_{i=1}^N k_{i}=K$.
The transmission procedure can be divided into two steps as follows.
\begin{enumerate}
\item In the first step, for the $i$th user requesting $W_{d_i}$, we transmit $W_{k,N(i-1)+j}$ ($k \neq d_i$ and $j=1,\ldots,N$), i.e., $(N-1)N$ subfiles in total are delivered to obtain $W_{d_i, N(i-1)+j}$ ($j=1,\ldots,N$) via coded operation.
\item In the second step, for the rest subfiles requested by users, we apply the following algorithm by firstly grouping the users requesting the same file and then applying coding strategy to reduce transmissions. The details are presented as follows.
\begin{enumerate}
\item If $W_{d_i}$ ($i=1,\ldots, K$) is solely requested by the $i$th user, all subfiles of $W_{d_i}$ can be completely received in Step 1). Hence the amount of remaining requests for $W_{d_i}$ is $0$.
\item For any $W_{i}$ requested by $k_i$ users ($k_i>1$), where each associated user requesting the residue $(k_i-1)N$ subfiles, we do
\begin{enumerate}
\item Initialization: list the users requesting $W_i$ in an ascending order with respect to their index. For simplicity, their index are correspondingly denoted by $K_l$ ($l=1, \ldots, k_i$). Observe that the exclusive subfiles obtained by user $K_l$ is $W_{i,N(K_l-1)+j}$ ($j=1,\ldots,N$) and they are requested by
    the other users in the same group. Set the initial value of the counter as $u=1$.
\item If $u=1$, deliver the $N$ coded subfiles, $W_{i,N(K_1-1)+j} \oplus W_{i,N(K_2-1)+j}$ ($j=1,\ldots,N$)
 and set $u \leftarrow u+1$.
\item If $u=m$ ($m<k_i-1$), deliver the $N$ coded subfiles, $W_{i,N(K_m-1)+j} \oplus W_{i,N(K_{m+1}-1)+j}$ ($j=1,\ldots,N$)
 to all users requesting $W_i$, set $u \leftarrow u+1$ and go to Step iv).
 \item If $u<k_i-1$ go to Step iii), otherwise terminate the delivery of subfiles of $W_i$.
\end{enumerate}
\end{enumerate}
\end{enumerate}

Note that in step 2), a) follows from two facts. The first is that the $i$th user obtains $W_{d_i, N(i-1)+j}$ ($j=1,\ldots,N$) via coded delivery. The second is that it receives directly
$W_{d_i, N(k-1)+j}$ ($k \neq i$ and $j=1,\ldots,N$) in the first step because they are delivered for other users for XORing. Therefore, the $i$th user can reconstruct the full file $W_{d_i}$ directly after Step 1).

Similarly for the case that $W_i$ is requested by more than one users ($k_i>1$) in b) of Step 2), the fact that each user requesting $W_i$ needs $(k_i-1)N$ follows also from two facts. The first is that it receives $N$ subfiles via coded delivery in Step 1). The second is that it directly receives $N(K-k_i)$ subfiles for the users requesting other files in Step 1). Therefore, only $NK-N-N(K-k_i)=N(k_i-1)$ subfiles is requested by each of the users requesting $W_i$.

In the following, we shall show that the sub-algorithm in b) in Step 2) can help all users requesting $W_i$ receive all the residue files.

Note that for user $K_m$ requesting $W_i$, it receives the subfile list
($W_{i,N(K_m-1)+j} \oplus W_{i,N(K_{m+1}-1)+j}$) ($m=1, \ldots, k_i-1$, $j=1,\ldots,N$).
It can firstly obtain $W_{i,N(K_{m-1}-1)+j}$ and $W_{i,N(K_{m+1}-1)+j}$ ($j=1,\ldots,N$) from the $m-1$th and the $m$ delivery of subfiles via XORing. It can then recursively obtain $W_{i,N(K_{m-k}-1)+j}$ ($k=2,\ldots,m-1$) and $W_{i,N(K_{m+k}-1)+j}$ ($k=2,\ldots,k_i-m$). Hence, user $K_m$ can obtain the complete file $W_i$. In a similar manner, we can verify that any other users in the same group requesting $W_i$ can receive the complete file $W_i$.

As $W_i$ is an arbitrary file in the server, we conclude that all users can obtain the requested file by our algorithm and in the following we shall derive the achievable rate for $M=1/K$ by applying the algorithm above. We first denote $C_i$ as the amount of subfiles delivered in Step i) and $n_{k_i}$ as the amount of the XORed version of subfiles delivered for $W_i$ in Step 2).

In Step 1), it is observed that the total amount of subfiles delivered is given by,
\begin{align}
C_1=(N-1)NK.
\end{align}

As designed in Step 2) for file $W_i$, the total amount of the remaining transmissions is
\begin{align}
n_{k_i}=(k_i-1)N. \label{eq:step_2_k_i}
\end{align}
Therefore, the total amount of subfiles delivered in the second step is
\begin{align}
C_2=&\sum_{i=1}^N n_{k_i}= \sum_{i=1}^N (k_i-1)N \label{eq:step_2_all_k_i_1}\\
= & \sum_{i=1}^N k_iN - N^2 = (K-N)N. \label{eq:step_2_all_k_i_2}
\end{align}
The total amount of subfile deliveries in these two steps is given by
\begin{align}
C_1+C_2=(N-1)NK+(K-N)N=(K-1)N^2.
\end{align}
The associated delivery rate therefore is
\begin{align}
R(1/K)=(K-1)N^2/NK=N-N/K>N-1
\end{align}
and we can claim that $(1/K,R(1/K))=(1/K,N(1-1/K)$ is an achievable cache-rate pair. In addition, regarding the trivial cache-rate pair $(0,N)$, for any $M \leq 1/K$, the rate pair $(M, N(1-M))$ is achievable by memory sharing. Theorem 2 is hence proved.


\subsection{Proof of Theorem 3 and Theorem 4}
Here we show that the achieved rate given in Theorem \ref{thm:3} and Theorem \ref{thm:4} for the scenario with $N \leq K$ and $M \leq 1/K$ coincides with the lower cut-set bound.

From \cite{niesen2014fundamental}, the cut-set lower bound is given by,
\begin{align}
R^*(M) \ge \max_{s\in\{ 1, \ldots, \min(N,K) \}}(s-\frac{s}{ \lfloor N/s \rfloor}M)
\end{align}

Therefore, with $M \le 1/K$, we obtain
\begin{align}
R^*(M) &\ge \max(1-\frac{M}{N}, \ldots, N-NM), \quad 0\leq M \leq \frac{1}{K} \label{eq:proof:T4_1}\\
& \ge N(1-M)  \label{eq:proof:T4_2}\\
& = R(M)    \label{eq:proof:T4_3}
\end{align}
where (\ref{eq:proof:T4_1}) follows directly from the cut-set bound and
(\ref{eq:proof:T4_2}) follows from the fact that $\max( \cdot )$ returns the maximum value of the elements in the brackets.
(\ref{eq:proof:T4_3}) follows directly from Theorem \ref{thm:1} and Theorem \ref{thm:2}.

From the above derivation, it is hence concluded that for the scenario $N \le K$ and $M \le 1/K$, the lower cut-set bound is achievable.
Theorem \ref{thm:3} and Theorem \ref{thm:4} are therefore verified.

%
%
%

\section{Conclusion}
In this work, we studied the caching problem when all users are with a small buffer size and the number of users is no less than the amount of files in the server. A novel coded caching scheme was proposed to achieve the cut-set bound rate for such a scenario.



\begin{thebibliography}{99}
\bibitem{dowdy1982file}
L. W. Dowdy and D. V. Foster, ``Comparative models of the file assignment problem,''
  \emph{ACM Comput. Surv.}, \hskip 0.5em plus
  0.5em minus 0.4em\relax vol. 14, no. 4, pp.
  287--313, Jun. 1982.

\bibitem{ammar1996multicast}
K. C. Almeroth and M. H. Ammar, ``The use of multicast delivery to provide a scalable and interactive video-on-demand service,''
  \emph{IEEE J. Sel. Areas Communi.}, \hskip 0.5em plus
  0.5em minus 0.4em\relax vol. 14, pp.
  1110--1122, Aug. 1996.

\bibitem{shahabuddin1996dynamic}
A. Dan, D. Sitaram and P. Shahabuddin, ``Dynamic batching policies for an on-demand video server,'' \emph{Multimedia Syst.}, \hskip 0.5em plus
  0.5em minus 0.4em\relax vol. 4, pp.
  112--121, June. 1996.

\bibitem{olotkin2001web}
A. Meyerson, K. Munagala and S. Plotkin, ``Web caching using access statistics,''
  \emph{Proc. 12th ACM-SIAM Symp. Discrete Algorithm (SODA'01)}, \hskip 0.5em plus
  0.5em minus 0.4em\relax pp.
  354--363, June. 2001.

\bibitem{swamy2008dataplacement}
I. Baev, R. Rajaraman and C. Swamy, ``Approximation algorithms for data placement problems,''
  \emph{SIAM J. Comput.}, \hskip 0.5em plus
  0.5em minus 0.4em\relax vol. 38, pp.
  1411--1429, July. 2008.

\bibitem{walid2010distributed}
S. Borst, V. Gupta, and A. Walid, ``Distributed caching algorithms for content distribution networks,''
  \emph{Proc. IEEE Int. Conf. Computer Communi. (INFOCOM'10)}, \hskip 0.5em plus
  0.5em minus 0.4em\relax pp.
  1478--1486, Mar. 2010.

\bibitem{kol2011demand}
Y. Birk and T. Kol, ``Coding on demand by an informed source (ISCOD) for efficient broadcast of different supplemental data to caching clients,''
  \emph{IEEE Trans. Inf. Theory}, \hskip 0.5em plus
  0.5em minus 0.4em\relax vol. 52, pp.
  2825--2830, Jun. 2006.

\bibitem{kol2011index}
Z. Bar-Yossef, Y. Birk, T. S. Jayram, and T. Kol, ``Index coding with side information,''
  \emph{IEEE Trans. Inf. Theory}, \hskip 0.5em plus
  0.5em minus 0.4em\relax vol. 57, pp.
  1479--1494, Mar. 2011.

\bibitem{ahlswede2000network}
R. Ahlswede, N. Cai, S. Li, and R. Yeung, ``Network information flow,''
  \emph{IEEE Trans. Inf. Theory}, \hskip 0.5em plus
  0.5em minus 0.4em\relax vol. 46, no. 4, pp.
  1204--1216, Apr. 2000.

\bibitem{niesen2014fundamental}
M. A. Maddah-Ali and U. Niesen, ``Fundamental limits of caching,''
  accepted by \emph{IEEE Trans. Inf. Theory}, \hskip 0.5em plus
  0.5em minus 0.4em\relax 2014.

\bibitem{niesen2014nonuniform}
U. Niesen and M. A. Maddah-Ali, ``Coded caching with nonuniform demands,''
  \emph{arXiv: 1308.0178 [cs.IT]}, \hskip 0.5em plus
  0.5em minus 0.4em\relax Aug. 2013.

\bibitem{niesen2014decentralized}
M. A. Maddah-Ali and U. Niesen, ``Decentralized caching attains order-optimal memory-rate tradeoff,''
  \emph{arXiv: 1301.5848 [cs.IT]}, \hskip 0.5em plus
  0.5em minus 0.4em\relax Jan. 2013.

\bibitem{molisch2013d2d}
M. Ji, G. Caire and A. Molisch, ``Fundamental limits of distributed caching in D2D wireless networks,''
  \emph{Proc. IEEE Inf. Theory Workshop (ITW'13)}, \hskip 0.5em plus
  0.5em minus 0.4em\relax Sep. 2013.

\bibitem{clancy2014secure}
A. Sengupta, R. Tandon and T. C. Clancy, ``Fundamental limits of caching with secure delivery,''
  \emph{arXiv: 1312.3961 [cs.IT]}, \hskip 0.5em plus
  0.5em minus 0.4em\relax Feb. 2014.

\bibitem{cover1991elements}
T. M. Cover and J. A. Thomas, \emph{Elements of Information Theory.}\hskip 0.5em plus
  0.5em minus 0.4em\relax
  Wiley, 1991.



\end{thebibliography}

\end{document}